\documentclass[aps,superscriptaddress,amsfonts,nofootinbib,twocolumn]{revtex4-1}
\usepackage{epsfig}
\usepackage{feynmp}
\usepackage{graphicx}
\usepackage{amssymb}
\usepackage{amsmath}
\usepackage{amsbsy}
\usepackage{graphicx}
\usepackage{amssymb}
\usepackage{amsmath}
\usepackage{amsbsy}
\usepackage{color}
\usepackage{dsfont}
\usepackage{upgreek}
\usepackage{soul}
\usepackage[normalem]{ulem}
\usepackage{color}
\usepackage{slashed}
\usepackage{amsbsy}
\usepackage{bm}
\usepackage{ulem}
\usepackage{chngpage}
\newcommand{\ee}{\end{equation}}
\newcommand{\bb}{\begin{equation}}
\newcommand{\eqb}{\begin{eqnarray}}
\newcommand{\eqf}{\end{eqnarray}}

\def\nablavec{\mbox{\boldmath$\nabla$}}

\usepackage{dsfont}
\usepackage{upgreek}

\definecolor{gre}{rgb}{0,0.8,0.3}

\usepackage{color}

\begin{document}
\title {\vspace{-2.05cm}
\hfill{\small{DESY 14-214}}\\[1.27cm]
Extracting  Hidden-Photon {Dark Matter} From a{n} LC-Circuit}

\author{Paola Arias}
\affiliation{Departmento de F\1sica, Universidad de Santiago de Chile, Casilla 307, Santiago, Chile}
\author{Ariel Arza}
\affiliation{Departmento de F\1sica, Universidad de Santiago de Chile, Casilla 307, Santiago, Chile}
\author{Babette D\"obrich}
\thanks{ {Now at CERN, CH-1211-Geneva 23}}
\affiliation{Deutsches Elektronen-Synchrotron DESY, 
Notkestr. 85, 22607 Hamburg, Germany}

\author{Jorge Gamboa}
\affiliation{Departmento de F\1sica, Universidad de Santiago de Chile, Casilla 307, Santiago, Chile}
\author{Fernando M\'endez}
\affiliation{Departmento de F\1sica, Universidad de Santiago de Chile, Casilla 307, Santiago, Chile}

\begin{abstract}We point out  that a cold dark  matter condensate made
  of gauge  bosons from an  extra hidden  U(1) sector -  dubbed hidden-
  photons  -   can  create  a  small,   oscillating  electric  density
  current.  Thus,  they could  also  be  searched  for in  the  recently
  proposed LC-circuit  setup conceived for  axion cold dark  matter search by
  Sikivie, Sullivan and Tanner.
{{We estimate  the sensitivity of this setup  for hidden-photon 
cold dark matter and we find it could cover {a sizable, so far  {unexplored}} parameter space}}.
\end{abstract}
\date{\today}
\maketitle
\section{Introduction}
Nowadays,  {direct} dark  matter  searches   are {mainly}  taking  two  alternative  and
complementary routes: one  of them aims to detect  {high-}{{mass}} candidates --
  so-called   Weakly  Interacting  Massive  Particles   (WIMPs)  --
exploiting scattering  experiments \cite{review1},  and the  other one
looks for light {{mass}} candidates -- so-called Weakly Interacting Slim
particles (WISPs) -- using precision experiments and strong magnetic
fields \cite{review2}.

Among  WISPs, the axion  is  a  prime candidate.  {It}  was
originally  proposed as  a mechanism  to solve  the strong  CP problem
\cite{Peccei:1977hh}.  {Soon} after  this  proposal,  it  was
realized that axions can be non-thermally produced
by a misalignment mechanism, making it a strong cold dark matter (CDM)
candidate   in  the   range   of  masses   $m_a  \lesssim   10^{-4}$eV
\cite{Preskill:1982cy}.

A  common  feature among  WISPs  is  their  weak coupling  to  the
Standard Model,  and the smallness  of their masses. 
{This is often {a}}  heritage from
the   high{-}energy   scale  at   which  their   underlying  symmetries
break. Many  indirect astrophysical  observations have  placed strong
constraints on these  particles \cite{Baker:2013zta}, but there is  still plenty of
parameter space  in which they could hide.
In particular, the parameter  space where
they can be CDM  remains still quite open.

The  WISPs relevant to this study are   
hidden sector  $U(1)$ gauge  bosons
\cite{holdom},    also    known     as    paraphotons,    or    hidden
photons.  Remarkably,  the same  non-thermal  mechanism  of axion  CDM
production   also  works to  produce  a condensate  of cold  hidden
photons  \cite{Nelson:2011sf,  wispycdm},  whose {viable} parameter space  
{spans}  a   wide
range and remains almost unconstrained by observations.

Consequently, experimental efforts  have  increased  in
lasts  years, and  several precision experiments  have been {and will be} set up, 
like  ADMX \cite{Wagner:2010mi},  ALPS
\cite{Ehret:2010mh}, CAST,  CROWS \cite{Betz:2013dza}, IAXO \cite{Irastorza:2013kda} (just to  name a
few)  {and help} to cover some of the unexplored parameter space.

Novel proposals, specially thought to reach the hinted cold dark matter  parameter space have emerged, 
such as  a dish antenna experiment
    {\cite{Horns:2012jf}}. 
    In this study we  want to {revisit} the proposal made  by Sikivie,
    Sullivan and Tanner \cite{sikivie}, {in which} they explore the particular form
taken by the Maxwell equations {{ if axion CDM is present}}.

This  new  setup  has  interesting  {{features}};  the  first  is  the
simplicity  of the  idea, namely  an LC-circuit  carrying an  electric
current generated  by CDM axions  in an external  magnetic field.
Secondly, the signal produced by  axions can be amplified by the
    circuit, making it detectable by  magnetic flux
detection  techniques.

The aim of  this letter is {{ to show that  hidden-photon CDM can
    also provide an oscillating electric  current, without the need of
    an external electromagnetic  field, which can act as  a source for
    the proposed experiment \cite{sikivie}.  Therefore, this setup can
    also hunt for these particles.}}
     {We note that LC circuits {{have been mentioned in \cite{Graham:2014sha} }}as hidden photons receivers, 
 however not adapted to the context of Dark Matter detection.}

{{The paper is organized as follows:  in section \ref{axion} we briefly review the
    operating  mechanism   of  the  LC  circuit   designed  to  detect
    axions. In section \ref{hidden} we show  how an oscillating current from hidden- photon 
CDM emerges  from the coupling of the  latter with photons,
    and we obtain the sensitivity of the experiment proposed in \cite{sikivie} for
    hidden photons. Finally in section \ref{conclu} we conclude.  }}

\section{Essentials of the axion search {with} an LC-circuit}\label{axion}

Let us recall the essentials of the proposal made in
\cite{sikivie}.  The idea exploits the fact that the coupling of
axions {{and photons}}

\bb
\mathcal L = - g\, a F_{\mu \nu} \tilde F^{\mu\nu}, \label{eq1}
\ee 
gives rise to a modified electrodynamics
 \bb \nablavec \times {\bf B}
- \frac{\partial {\bf E}}{\partial t} =  - g \,{\bf B} \frac{d a}{d t}
+ {\bf J}_{\text{ext}} 
\label{01} 
\ee 
where  $g$ is  the coupling constant  between axions  and photons,
${\bf J_{\rm ext}}$ is an external source and a homogeneous axion field
is assumed  {\it i.e.}  $a=a(t)$ and  {{therefore spatial derivatives
  of this field are   neglected. This is approximately valid for axion DM .}} 

Note that  eq.~(\ref{01}) contains the effective  displacement current
{{$ {\bf{j_a}}=-g \,{\dot  a} \,{ \bf B_0} $}}  which {{emerges}} when
an {{{external magnetic field ${\bf B}_0$ is turned on and,}} as a
  consequence  the current  ${\bf  j}_a$ becomes  a  source for  {{a}}
  magnetic field $ {\bf B} _a $ through the equation $\nablavec \times
  {\bf B}_a = {\bf j}_a$.
  
Thus, the idea is to insert part of an LC-circuit into a region with an external magnetic field.
Fig.~(\ref{fig:1}) mimics the setup of \cite{sikivie}, 
where {{the}} external magnetic field $\bf{B}_0$ around the passive part of 
the circuit is omitted for hidden photons search (see below). 

\begin{figure}
\centering
\includegraphics[scale=0.7]{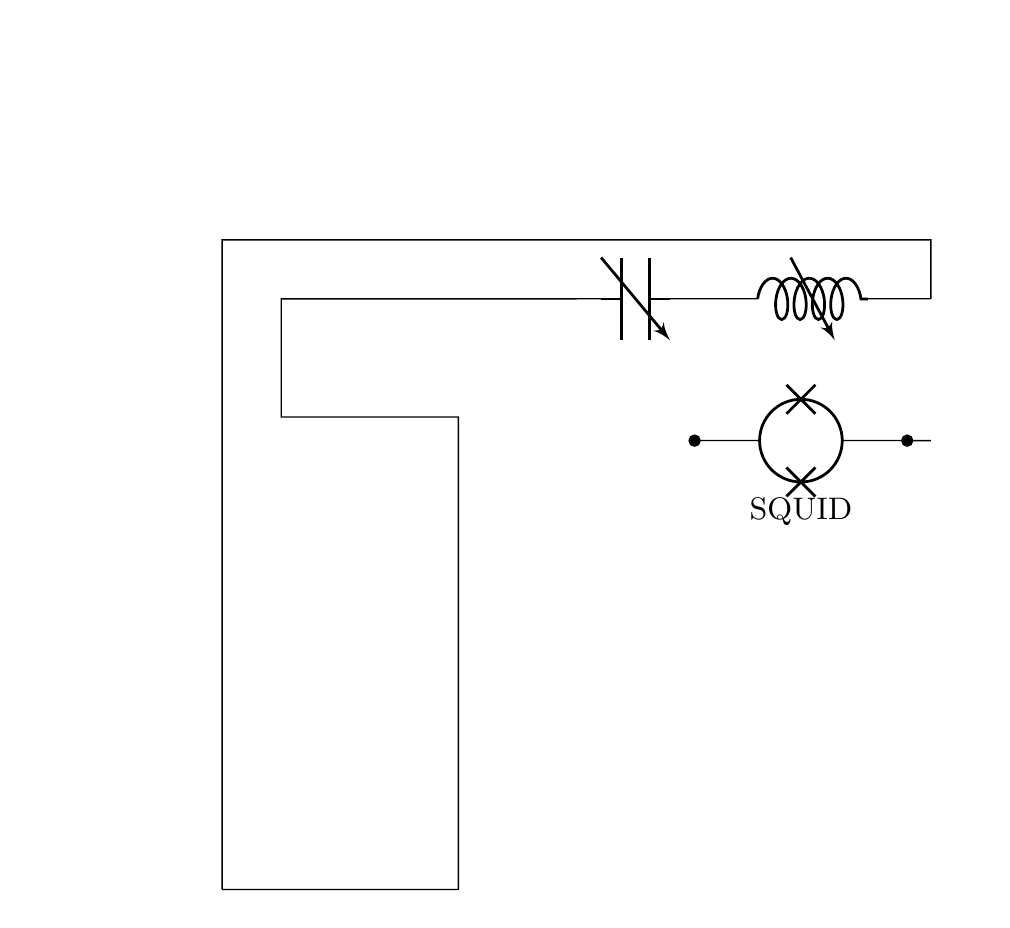}
\caption{ Sketch
of the experiment. In contrast to the original Sikivie-Sullivan-Tanner
setup  \cite{sikivie}, no external magnet is needed.
As the hidden-photon DM might have a net polarization,  orthogonal circuits
{could} be foreseen to minimize the effect of $\kappa$ 
(see text for details).  
}
\label{fig:1}
\end{figure}

The modification  (\ref {01}) implies that for the setup \cite{sikivie} in the presence of axion CDM  
an electric current will flow through the 
{LC circuit}, eventually in the resonance regime for $1/\sqrt{LC}\approx m_a$.

The  important  thing to  note,  however, is  the
following;  once the  electric  current is  produced {{in}} the  circuit,
the magnetic flux in the coil measured  by a magnetometer, here a SQUID, is  related to
the dark matter axion properties.

Indeed,
the amplitude  of the magnetic  field $B_d$ in the coil that  will be
detected by the  SQUID  is given by 
\bb
B_d \simeq \frac{N_d Q}{2r_d L} V_m g B_0 \sqrt{2\rho_{\rm DM}},
\label{Baxion}
\ee  where $N_d,r_d, V_m$ are parameters
  of the  device keeping the notation
  of \cite{sikivie}:
{$N_d$ is the number of turns and
  $r_d$ is radius of the small coil on the right hand side of Fig. \ref{fig:1},
  $V_m$ is a parameter with dimensions of volume, which appears in
  the integration of the magnetic flux in the circuit, see below.}
  $L$ is the  inductance of the entire  system, $Q$ is
  the quality factor {{of the circuit}}, $B_0$ is
  the magnitude of the external  magnetic field and $\rho_{\rm DM}$ is the
  dark matter energy density of the axion field.

{{In their proposal \cite{sikivie}, the authors have assumed a superconducting circuit, working at a temperature of the order of $T \sim 0.5$mK, a SQUID 
sensitivity of the order of $10^{-15}$T,  a quality factor of the circuit  $Q=10^4$, and two possible magnets: the ADMX magnet or CMS magnet (see details in \cite{sikivie}).

\section{Connection with hidden  photons}\label{hidden}
In this section we will argue that an oscillating current also emerges
if the CDM content is composed of hidden  photons.

{In the hidden photon model we are interested in, the}
dominant interaction between hidden photons and our visible sector
{is} {{via}}  a kinetic mixing term.
At low energies the effective Lagrangian is given by
\bb  \mathcal{L}= -\frac{1}4 F_{\mu \nu} F^{\mu
  \nu} -\frac{1}4X  _{\mu \nu}  X^{\mu \nu} -\frac{\chi}2  F_{\mu \nu}
X^{\mu \nu} +\frac{m_{\gamma'}^2}2 X_\mu X^\mu.
\ee 

{Here}  $F_{\mu\nu} $  is the  field strength  associated with  photons
($A_\mu$) and $X_{\mu \nu}$ the analogue for hidden photons ($X_\mu$).
The $\chi$ parametrizes the strength of the coupling between both, and
is predicted to be small \cite{Dienes:1996zr}.  The mass of the hidden photon {$m_{\gamma'}$}
 can be generated via a hidden-Higgs mechanism, or a
    St\"uckelberg mechanism.  

In an analogue way to axions, hidden-photon CDM can be considered as a homogeneous field
in    space,   given    by    ${\bf{    X}}(t)={\bf{   X}}_{\rm    DM}
e^{-im_{\gamma'}t}$, where  $\bf{X}_{\rm DM}$ is  the
DM  vector, and  due to   its  vector nature,  a cold
condensate  of  hidden photons   {can introduce}  a  preferred direction  in
space.

{Effectively}  a tiny  fraction of {its} energy  is invested  in an
ordinary oscillatory electric field \cite{Horns:2012jf}, given by
\bb
{\bf E}_{\rm DM}= \chi\, m_{\gamma'} {\bf X}_{\rm DM} e^{-im_{\gamma'}
  t}. 
\ee
Such  an electric  field  will  create  a  displacement
current, {{oscillating  at the same  frequency as the  electric field,
    $\nu=   0.24$~GHz   $
    (m_{\gamma'}/\mu\rm{eV})$  \cite{Horns:2012jf} }},
    given by 
\bb
{\bf{J}}_{\rm HP}= -\frac{\partial  {\bf{E}}_{\rm DM}} {\partial t}.
\label{currenthp}
\ee
The amplitude, analogously to axions, is related
to  the CDM  local density  by noting  that the  stored energy  in the
condensate is
\bb
\rho_{\rm      DM}      \sim     300\,      \frac{\rm{MeV}}{\rm{cm}^3}
=\frac{m_{\gamma'}^2}2 \langle |{\bf{X_{\rm DM}}}|^2\rangle. 
\ee
Thus, the corresponding magnitude of the current density  obtained is
\bb
{{
|{\bf{J}}_{\rm HP}|= \chi m_{\gamma'} \sqrt{2\rho_{\rm DM}}.
}}
\ee
{{This current will generate - in principle -  oscillating {{electromagnetic fields.  Nevertheless, assuming the experiment is enclosed {{\footnote{{{With enclosed we mean  the electric field, ${\bf{E}}_{\rm HP}$, is set to zero at a  boundary.}}}}} in  a region of characteristic dimension smaller than $ m_{\gamma'}^{-1}$, we can work in the magneto-quasistatic limit (also assumed in \cite{sikivie}). Following this approximation, the magnetic field created by the displacement current is just $\nablavec \times {\bf{B}}_{\rm HP}=
    {\bf{J}}_{\rm HP}$,  and the electric field induced is obtained from $\nabla \times {\bf{E}}_{\rm HP}=-\partial {\bf{B}}_{\rm HP}/ \partial t$.  The latter is suppressed {{inside the enclosed region}} in comparison with the magnetic field by $|{\bf{E}}_{\rm HP}|= m_{\gamma'}r |{\bf{B}}_{\rm HP}|$, where $r$ is the radial distance  in cylindrical coordinates, with the symmetry axis parallel to the
    direction of the superconducting wire of fig.~(\ref{fig:1}). Therefore, in the following,  we assume the induced electric field (${\bf E}_{\rm HP}$) does not interfere significantly with the small electronic devices of the circuit.}}


Note that in the present   case of hidden photons, the
displacement current, {see}  eq.~(\ref{currenthp}), has the {{same}} direction {as} the CDM condensate,} if effectively it has a preferred direction (see below)}.   In the  case  of  axions,
instead,  the  current  density  has the  direction  of  the  external
magnetic field,  ${\bf{B}}_0$.

{{The component of the current }}${\bf J}_{\rm HP}$ which
  is  parallel  {to the superconducting wire direction}  will
  contribute  to create  the components  of field  ${\bf{B}}_{\rm HP}$
  which will be  responsible for a non zero magnetic  flux through the
  same part of the circuit. Namely 
\bb 
|{\bf J}_{ {\rm HP}\parallel}|=|\bf{J}_{\rm HP} \cos \theta|{{=|{\bf{J}}_{\rm{HP}}|\kappa}},
\ee 
where $\theta$ is the angle  between }  {the wire} and  the direction of the  current generated
by the hidden-photon condensate, {{and $\kappa=|\cos\theta|$.}}

This last analysis yields to the two possible scenarios:
\begin{itemize}
\item[i)] The condensate of DM points in a preferred direction in space, $\hat n$.
\item[ii)] The condensate of DM is randomly oriented in space 
\end{itemize}

\begin{figure}[t]
\vspace{0.5cm}
\centering
\includegraphics[scale=0.45]{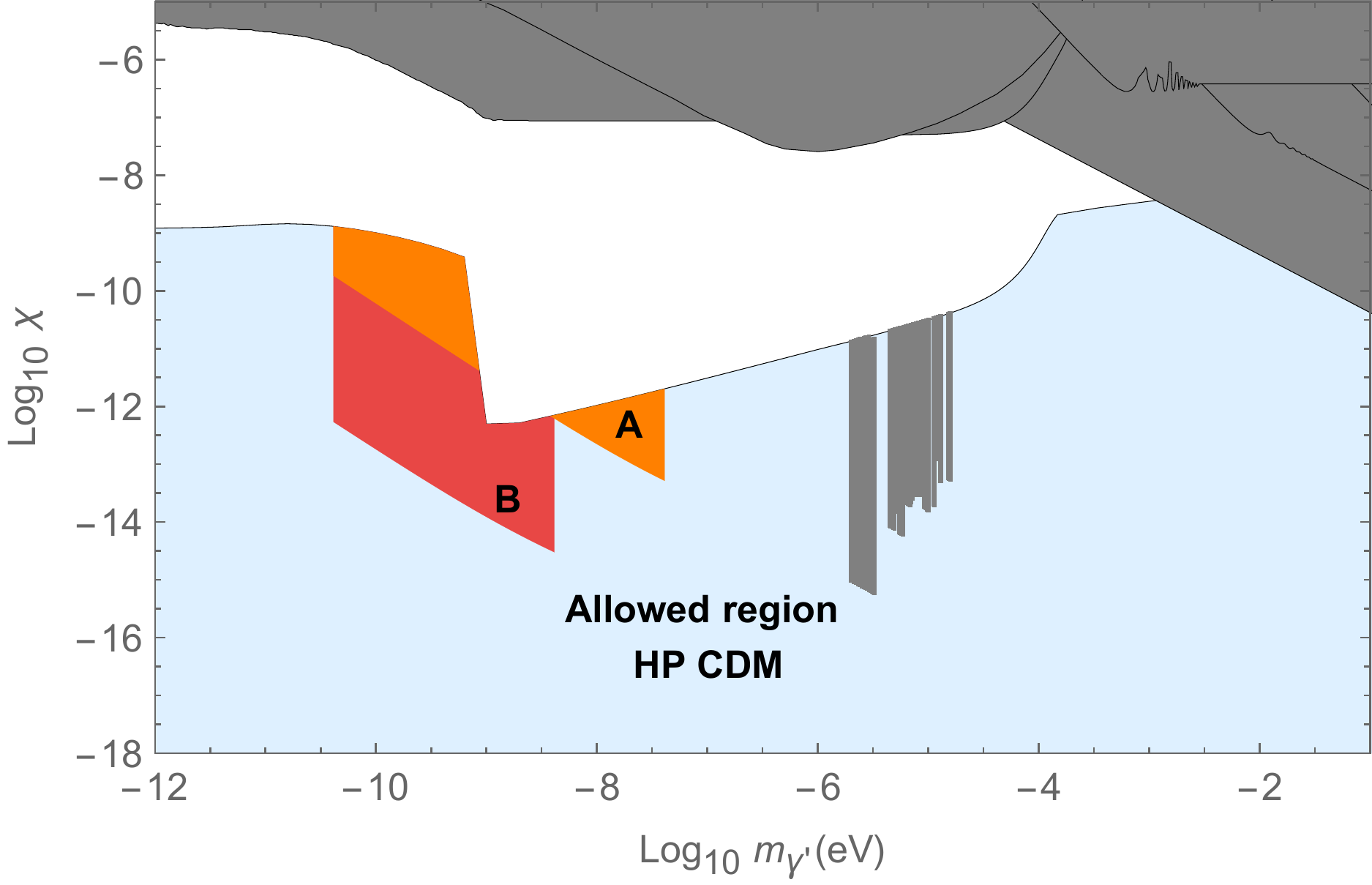}
\caption{Parameter space scanned by the LC circuit proposed in \cite{sikivie} 
for hidden-photon cold dark matter (see the text for details).
\label{fig:2}
} 
\end{figure}

{{In the}} first  case  each direction $\hat  n$ is equally
probable; {{a}}   conservative {{choice for}}  ${{\kappa }}$
{{would be, for instance, }} that  its real value  is bigger  with a 95$\%$  confidence level,
thus $\kappa= 0.05$.  

In the second case {{we consider}} the average among all possible angles
{{and then}} ${{\kappa}}=0.5$.

{{{By the other hand, {if scenario i) is realized in nature, the lab frame's movement with respect to
the rest frame of the DM will {likely} yield {{a non-constant }}$ \theta$.
{In fact, this signal modulation could help to track down the
Dark Matter nature of the signal.}
{Also,} additional setups in which {an} LC loop is oriented orthogonally
to the primary setup would allow to probe the parameter space with
{{$\kappa\simeq 1$}}.}}}

{{Assuming the superconducting wire is oriented in the $\hat z$ direction, the magnetic field induced by the HP-DM is given by
\bb 
{\bf{B}}_{\rm HP}=  -\frac{1}2 \chi m_{\gamma'}^2 |{\bf{X}}_{\rm DM}| e^{-im_{\gamma'}t} r \kappa \,\, \hat \phi.
\ee

}}

At this point, let us note that the current generated in the circuit is given by $I= \Phi /L$, where $\Phi$ is the magnetic flux of the field generated by the dark matter, and $L$ is the inductance of the circuit. In the case of axions, the magnetic flux is
\bb
\Phi_a= -gB_0 \dot a V_m,
\ee
where   $V_m= \frac{1}{4} l_m^2 r^2_m$ and $l_m$ and
$r_m$ are the length and width of the circuit loop immersed in the
field and thus are limited by the size of the magnet bore.

Since hidden-photon DM does not need an external electromagnetic field to induce a current in the circuit, the latter one is also given by $I=\Phi_{\rm HP}/L$, {{and}} the magnetic flux {{is}}

\bb
\Phi_{\rm HP}=-\kappa \chi m_{\gamma'}{{^2}} |{\bf{X}}_{\rm DM}| {{e^{-i m_{\gamma'}t}}}V'_m,
\ee 
where $V'_m$ includes now also the part of the
loop not immersed in the external field.
{We will assume in the following  $V_m' \simeq V_m$
 to be realized in the setups of \cite{sikivie}.
 In principle, the additional contribution
 incorporated in $V'_m$ can be used to further enhance the signal,
however details depend on the exact geometry of the experimental setup:
As $V_m$,  $V'_m$ can be obtained from an integral along the
loop-area transversal to ${\bf{ B}}_{\rm HP}$. 
}

{As pointed out in \cite{Hong:2014vua}, the geometry
of the outer volume, {\it e.g.} a cavity enclosing the LC circuit,
can have {positive} impact on the overall quality factor. 
{In the following, however we stick to the original setting of  \cite{sikivie}}.}

From the hidden photon point of view the magnetic flux measured in the coil of the Sikivie et al setup \cite{sikivie} is given by  
\bb
B_{\rm {{detected}}} = \frac{N_d Q}{2r_d L} V_m \chi m_{\gamma'} \sqrt{2\rho_{\rm DM}} \,\kappa.
\ee

{{To get a 
{sensitivity estimate} for hidden photons we consider the isothermal halo model \cite{MST},
where the local dark matter density is $\rho_{DM}=0.3$GeV/cm$^3$. 
The energy dispersion $\delta E \sim 10^{-6} m_{\gamma'}$, 
is then bigger than the one considered in \cite{sikivie}, leading to a reduced coherence time. 
The latter translates into a different  magnetometer's sensitivity, now given by
\bb
\delta B=10^{-16}\, \rm{T}\,\left(\rm{Hz}\right)^{-1/2} \left(t_c t\right)^{-1/4} \ ,
\ee
{see also \cite{Budker:2013hfa} for a {detailed} 
discussion of the sensitivity scaling.}
 {If we} consider the experiment to run parasitically to the search for axion DM,  the measurement time is $t=10^3$s, and the coherence time is $t_c=0.16 \,\rm{s}\left(\rm{MHz}/\nu\right)$.
}}
{{In order to compare with}} \cite{sikivie} we have
      chosen a signal-to-noise ratio of 5.}
{{In fig.~(\ref{fig:2}) we show the }}parameter  space  {{that could be}}  scanned  by   
the  LC  circuit   proposed  in
      \cite{sikivie} for  hidden-photon  cold dark matter. The light
    blue area corresponds to the  allowed parameter space of  hidden-photon
     cold dark matter \cite{wispycdm}. {{The orange
    region corresponds to the sensitivity of the experiment  running with the ADMX magnet (A),
    $V_m=0.023$~m$^3$, while the red region assume  the
    setup working with  the CMS  magnet (B), $V_m=29.25$~m$^3$   (for details of these two
    magnets see  \cite{sikivie}).
    Gray  areas correspond  to previously
    excluded regions.   {{We have  considered scenario ii) and used $\kappa=0.5$. }}
  {We emphasize again that the HP setup could in principle
  profit from the fact that is not necessary to magnetize the volume.
  However, we stick to the estimates above, because cryogenics and shielding
  at such values are demanding by themselves.
  This then sets the lower scanable frequency, whereas we take
  the high-frequency cutoff through stray capacitance as in  \cite{sikivie}.}

\section{Conclusions}\label{conclu}

{{If cold dark matter is made of hidden photons, it can source electric and magnetic fields, since  modifies Maxwell equations in a similar way as axions (and axion-like particles) do.}}  We  have  pointed   out  that  the  proposal  of
    Sikivie-Sullivan-Tanner {to search
for axionic CDM with an LC circuit}  could also  be  used to  test the  hidden-
    photons as dark matter candidate. {{The projected sensitivity of the experiment for hidden photons can cover unconstrained parameter space, as shown in fig.~(2).}}
  
  {We emphasize again that our proposal
    has the \lq advantage\rq  $\,$ over  \cite{sikivie} that no strong external magnet is needed,
    but rather just  cryogenic volume and appropriate electronics.
    
    We believe that this makes our proposal attractive to a larger 
    group of experimentalists without access to
    strong magnets. Given the huge discovery potential for hidden-photon
    cold Dark Matter, also a dedicated search that runs non-parasitically to
    the axionic equivalent constitutes a worthwhile fundamental physics experiment.}

We would like to  thank Joerg Jaeckel and Pierre Sikivie for  comments. This work was supported by FONDECYT/Chile grants   and ACT1102 (P.A.), 1130020 (J.G.), 1140243  (F.M.) and  Conicyt-21120890 (A. A.).

\vskip 0.5cm
\noindent{\underline{\bf Note added}}

{One  week after the present paper was submitted to {{a}}rXiv, an {{ extensive
proposal for a DM search for Hidden Photons was put forward}} in
 \cite{Chaudhuri:2014dla}. Even though their detection technique of the hidden photon DM condensate 
 is  different to the one considered here \cite{sikivie},  they {detailed on}
 the important point that the 
 setup {{needs}} to be shielded. 

{{We agree a shielding is needed and in fact is necessary to have it in order to  work in the magneto-quasistatic limit as implied in our calculation.}}

\end{document}